\documentclass[12pt]{article}
  \usepackage{amsthm,amsfonts}
  \usepackage{amsmath}
\newcommand{\bea}   {\begin{eqnarray}}
\newcommand{\eea}   {\end{eqnarray}}
\usepackage{subeqnarray}
\usepackage[brazilian]{babel}
\usepackage[latin1]{inputenc}
\begin{document}
\renewcommand{\thefootnote}{\fnsymbol{footnote}}
\date{}

\begin{center}
\Large{\bf Dimensionalidade do espa\c{c}o ou da extens\~{a}o:}\\ 
\Large{\bf nota sobre a contribui\c{c}\~{a}o do jovem Kant}\footnote{Publicado em \textit{Tempo Brasileiro} v.~189/190, p. 37-47 (2012).} 
\end{center}

\vspace*{0.4cm}

\begin{center}
\textit{Francisco Caruso$^{1,2}$ \& Zulena dos Santos Silva}$^2$\\
\vspace*{0.5cm}

${}^1$ Centro Brasileiro de Pesquisas F\'{\i}sicas\\
${}^2$  Programa de Hist\'{o}ria das Ci\^{e}ncias e das T\'{e}cnicas e Epistemologia da \\
Universidade Federal do Rio de Janeiro\\
\vspace*{0.2cm}
\end{center}

\vspace*{3mm}
\noindent

\vspace*{3mm}
\noindent

\vspace*{1.5cm}

\section{Introduction}
\setcounter{footnote}{0}
\renewcommand{\thefootnote}{\arabic{footnote}}

Neste artigo, apresentam-se novos coment\'{a}rios ao texto \textit{On Kant's First Insight Into The Problem of Space Dimensionality and Its Physical Foundations}\footnote{Francisco Caruso \& Roberto Moreira Xavier de Ara\'{u}jo, On Kant's first insight into the problem of space dimensionality and its physical foundations, arXiv:0907.3531, submetido \`{a} \textit{Kant Studien}.}  e, em particular, a como o espa\c{c}o \'{e} concebido no primeiro escrito de Kant. Em seu trabalho, datado de 1747, o jovem Kant prop\~{o}e-se a discutir como se pode justificar a tridimensionalidade do espa\c{c}o.\footnote{I. Kant, \textit{Gedanken von der wahren Sch\"{a}tzung der lebendigen Kr\"{a}fte und Beurtheilung der Beweise, deren sich Herr von Leibniz und andere Mechaniker in dieser Streitsache bedient haben, nebst einigen vorhergehenden Betrachtungen, welche die Kraft der K\"{o}rper \"{u}berhaupt betreffen}, K\"{o}nigsberg, 1747; reprinted in: \textit{Kant Werke}, Band 1, Vorkritische Schriften, Wissenschaftliche Buchgesellschaft Darmstadt, 1983. Tradu\c{c}\~{a}o para o ingl\^{e}s de parte deste trabalho foi feita por J. Handyside e publicada no livro \textit{Kant's inaugural dissertation and the early writings on space}, Chicago, Open Court, 1929, reprinted by Hyperion Press, 1979.}  Apesar de explicitamente antecipar que ir\'{a} demonstrar isto no decorrer do texto, na verdade, limita-se a relacionar a dimensionalidade da extens\~{a}o \`{a} lei da gravita\c{c}\~{a}o de Newton. De qualquer forma, esta especula\c{c}\~{a}o de Kant foi al\'{e}m da Filosofia e teve grande impacto na discuss\~{a}o moderna acerca da dimensionalidade do espa\c{c}o.\footnote{F. Caruso \& R. Moreira Xavier, \textit{Causa Efficiens versus Causa Formalis}: Origens da Discuss\~{a}o Moderna sobre a Dimensionalidade do Espa\c{c}o, \textit{Cadernos de Hist\'{o}ria e Filosofia da Ci\^{e}ncia}, Campinas, ser.~3, v. 4 (2), p.~43-64 (1994).}  De fato, do ponto de vista da F\'{\i}sica, sua conjectura pode ser compreendida no \^{a}mbito da formula\c{c}\~{a}o de campos cl\'{a}ssica da Teoria Newtoniana da Gravita\c{c}\~{a}o. Entretanto, alguns pontos filos\'{o}ficos que levaram o jovem Kant a n\~{a}o ter \^{e}xito em provar a natureza tridimensional do espa\c{c}o ainda merecem aten\c{c}\~{a}o. \'{E} sobre isto que trata este artigo.

\newpage
\section{Breve exposi\c{c}\~{a}o dos pressupostos nas ideias do jovem Kant sobre o espa\c{c}o}

Envolvidos na argumenta\c{c}\~{a}o mecanicista de Kant, constru\'{\i}da com o objetivo de explicar porqu\^{e} s\~{a}o tr\^{e}s as dimens\~{o}es do espa\c{c}o, est\~{a}o no\c{c}\~{o}es de ``causa'' e ``efeito'', subtendidas como o sentido dos acontecimentos da natureza, e as de ``for\c{c}a'', ``subst\^{a}ncia'', ``extens\~{a}o'' e ``espa\c{c}o'', as quais se conjugam na seguinte concep\c{c}\~{a}o encontrada no in\'{\i}cio de sua formula\c{c}\~{a}o do problema: ``Se as subst\^{a}ncias n\~{a}o tiverem uma for\c{c}a a partir da qual possam atuar fora de si pr\'{o}prias, n\~{a}o haveria extens\~{a}o e consequentemente nenhum espa\c{c}o.''.\footnote{Kant, 1979, p.~7-8.}  Em outras palavras, sua argumenta\c{c}\~{a}o baseia-se em uma sequ\^{e}ncia cognitiva que parte da for\c{c}a exercida sobre os corpos ou subst\^{a}ncias, a partir do que se fazem revelar a extens\~{a}o e, por conseguinte, o espa\c{c}o.

Neste ponto, pode-se notar que Kant parece associar espa\c{c}o \`{a} percep\c{c}\~{a}o, assim: inicialmente concebe o espa\c{c}o como dependente da extens\~{a}o, e esta como resultante do movimento ou a\c{c}\~{a}o de um corpo que comporta algo substancial sobre outro corpo (ou sobre a alma), a\c{c}\~{a}o esta sintetizada por uma for\c{c}a; quer dizer, s\'{o} h\'{a} o espa\c{c}o e a extens\~{a}o na medida em que haja uma intera\c{c}\~{a}o entre corpos oriunda de uma for\c{c}a; \'{e} o resultado desta intera\c{c}\~{a}o o que se percebe, e n\~{a}o propriamente o espa\c{c}o ou a extens\~{a}o, pois estes n\~{a}o preexistem \`{a} for\c{c}a como causa \'{u}ltima, por assim dizer, e s\~{a}o identific\'{a}veis na disposi\c{c}\~{a}o dos corpos ou subst\^{a}ncias entre si, o que \'{e} operado pela for\c{c}a. Mas tudo isso \'{e} teorizado, o espa\c{c}o talvez n\~{a}o seja percebido, mas depreendido do que de fato \'{e} fenom\^{e}nico: movimentos e corpos em intera\c{c}\~{a}o. Ocorre uma explica\c{c}\~{a}o f\'{\i}sica cr\'{\i}vel para nossa percep\c{c}\~{a}o dos acontecimentos. 

Vale tamb\'{e}m recordar que Kant est\'{a} compartilhando a concep\c{c}\~{a}o de Leibniz quanto ao espa\c{c}o ser relacional, e n\~{a}o a vis\~{a}o de Newton de espa\c{c}o absoluto, como recept\'{a}culo da mat\'{e}ria e dos acontecimentos. Assim, o engenhoso argumento de Kant conjuga ideias newtonianas sobre mat\'{e}ria, for\c{c}a gravitacional e as de Leibniz de espa\c{c}o relacional e vis viva. Cabe notar, que do ponto de vista da F\'{\i}sica, ele faz uma grande confus\~{a}o, ao associar a vis viva (que hoje se sabe tratar de algo como energia cin\'{e}tica) \`{a} for\c{c}a gravitacional. No tocante \`{a}s ideias de Leibniz, admite Kant que os corpos possuem uma for\c{c}a essencial, inerente \`{a} mat\'{e}ria e que antecede a pr\'{o}pria extens\~{a}o. Assim, a intera\c{c}\~{a}o de subst\^{a}ncias f\'{\i}sicas \'{e} depreendida da seguinte sequ\^{e}ncia causal: for\c{c}a, a causa primeira, rela\c{c}\~{a}o entre corpos, extens\~{a}o ou ordem ou disposi\c{c}\~{a}o astron\^{o}mica dos corpos, e espa\c{c}o, por fim. 

\section{A argumenta\c{c}\~{a}o de Kant acerca da pretensa dimensionalidade do espa\c{c}o}

Vejamos agora como se desenvolve o argumento de Kant sobre a tridimensionalidade do espa\c{c}o, ou melhor, da extens\~{a}o, a partir da lei da gravita\c{c}\~{a}o de Newton. Segundo Caruso e Moreira,\footnote{Francisco Caruso \& Roberto Moreira Xavier de Ara\'{u}jo, \textit{op.cit.}.}  o racioc\'{\i}nio de Kant desdobra-se nos seguintes passos:\\

 $1^\circ$) a aceita\c{c}\~{a}o de que existe uma for\c{c}a inerente \`{a}s subst\^{a}ncias, sem a qual n\~{a}o haveriam extens\~{a}o e rela\c{c}\~{a}o -- ou seja, sem mat\'{e}ria n\~{a}o se pode perceber a extens\~{a}o ou o espa\c{c}o. Logo, este n\~{a}o \'{e} apreens\'{\i}vel por si, apenas pode ser cogitado logicamente e da\'{\i} at\'{e} pode ser consequentemente concebido como ``espa\c{c}o absoluto'', mas, se assim, pende para o dom\'{\i}nio metaf\'{\i}sico; necess\'{a}rio, portanto, este primeiro passo da argumenta\c{c}\~{a}o n\~{a}o basear-se na ideia de Newton do ``espa\c{c}o absoluto'', mas da lei f\'{\i}sica da for\c{c}a;\\

 $2^\circ$) a hip\'{o}tese de que tal for\c{c}a \'{e} necess\'{a}ria para estabelecer a rela\c{c}\~{a}o entre os corpos, para a ordem ou disposi\c{c}\~{a}o astron\^{o}mica, e sem tal ordem o espa\c{c}o n\~{a}o existiria. Vemos neste pressuposto que o conceito de for\c{c}a assume um papel fundamental, como ess\^{e}ncia da mat\'{e}ria e de sua extens\~{a}o, na explana\c{c}\~{a}o de Kant, o que podemos conferir nas suas pr\'{o}prias palavras:\footnote{Kant, \textit{op. cit.}, p.~11.}

\begin{quotation}
\noindent \textit{``Como tudo que deve ser encontrado entre as qualidades de uma coisa deve ser capaz de ser derivado daquilo que cont\'{e}m em si a base mais complete da pr\'{o}pria coisa, as qualidades da extens\~{a}o, e subsequentemente sua natureza tridimensional, deve basear-se nas qualidades da for\c{c}a que as subst\^{a}ncias possuem com rela\c{c}\~{a}o \`{a}s coisas com as quais est\~{a}o conectadas.''}
\end{quotation}

E sobre a natureza dessa for\c{c}a fundamental, temos a seguinte cita\c{c}\~{a}o:\footnote{Kant, \textit{op. cit.}, p.~12.}

\begin{quotation}
\noindent \textit{``A for\c{c}a, pela qual uma subst\^{a}ncia age em uni\~{a}o a outras, n\~{a}o pode ser pensada deixando-se de lado uma determinada lei que se revela no modo de sua a\c{c}\~{a}o. Uma vez que o car\'{a}ter dessas leis de acordo com as quais toda uma cole\c{c}\~{a}o de subst\^{a}ncias (ou seja, espa\c{c}o) \'{e} medida, em outras palavras, a dimens\~{a}o da extens\~{a}o, ser\'{a} igualmente devida \`{a}s leis segundo \`{a}s quais as subst\^{a}ncias por meio de suas for\c{c}as essenciais tentam se conectar.''}
\end{quotation}

Os autores notam que tal compreens\~{a}o -- da for\c{c}a atuando sobre a subst\^{a}ncia, da\'{\i} provocando intera\c{c}\~{o}es do que, por sua vez, resulta a extens\~{a}o, uma vez que a subst\^{a}ncia age para fora de si, e da\'{\i} fazendo surgir o espa\c{c}o -- revela o espa\c{c}o como ``subsidi\'{a}rio'', intelig\'{\i}vel a partir das rela\c{c}\~{o}es entre as subst\^{a}ncias, compat\'{\i}vel com a vis\~{a}o de espa\c{c}o relacional de Leibniz.
Apenas cogit\'{a}vel, e n\~{a}o percept\'{\i}vel, o espa\c{c}o n\~{a}o \'{e}, portanto, tido como um fen\^{o}meno f\'{\i}sico. Kant, desse modo, se distingue de Newton ao n\~{a}o considerar, como este o fez, o espa\c{c}o absoluto; o programa newtoniano vale para Kant na medida em que oferece uma lei f\'{\i}sica capaz de registrar matematicamente os fen\^{o}menos gravitacionais da natureza, mas quanto \`{a} dimensionalidade do espa\c{c}o isso n\~{a}o se d\'{a} de modo claro. Parece-nos que o jovem Kant tenha notado isso, visto que, por mais vezes e em trechos importantes de sua argumenta\c{c}\~{a}o, usa o termo ``extens\~{a}o'' e n\~{a}o ``espa\c{c}o''. 

\section{Extens\~{a}o ou espa\c{c}o?}

Para Caruso \& Moreira,\footnote{\textit{Op. cit.}.}  isso indica o quanto ``Kant n\~{a}o compartilha com Galileu a ideia da possibilidade de geometriza\c{c}\~{a}o da F\'{\i}sica''. Quanto a este ponto, gostar\'{\i}amos de propor aqui uma interpreta\c{c}\~{a}o diferente. O jovem Kant concebe a realidade f\'{\i}sica a partir de uma sequ\^{e}ncia que tem in\'{\i}cio na for\c{c}a respons\'{a}vel pelas intera\c{c}\~{o}es entre corpos, do que decorre a extens\~{a}o, sequ\^{e}ncia esta pass\'{\i}vel de medi\c{c}\~{a}o, o que ele argumenta partir da lei gravitacional, a qual \'{e} matematiz\'{a}vel.  Pode-se imaginar que ele acaba deixando de fora da argumenta\c{c}\~{a}o o espa\c{c}o, e acaba se detendo na extens\~{a}o, porque aquele n\~{a}o se lhe mostra matematiz\'{a}vel, com o que n\~{a}o se poderia registrar sua condi\c{c}\~{a}o f\'{\i}sica, percept\'{\i}vel. N\~{a}o sendo matematiz\'{a}vel, o espa\c{c}o lhe parece ter uma natureza metaf\'{\i}sica, n\~{a}o sendo, portanto, um fen\^{o}meno f\'{\i}sico. Kant ent\~{a}o n\~{a}o conturba o projeto de Galileu de geometriza\c{c}\~{a}o da F\'{\i}sica; ao contr\'{a}rio, lhe d\'{a} sustenta\c{c}\~{a}o na medida em que parece supor que o crit\'{e}rio para uma ideia ou conceito ser correspondente a um fen\^{o}meno f\'{\i}sico -- e n\~{a}o metaf\'{\i}sico -- reside na possibilidade de matematiza\c{c}\~{a}o do mesmo.

De fato, j\'{a} neste texto pr\'{e}-cr\'{\i}tico que estamos analisando, Kant relegaria o espa\c{c}o para o dom\'{\i}nio da Geometria, uma vez que n\~{a}o se mostra como dado sens\'{\i}vel, podendo apenas ser cogitado por formula\c{c}\~{o}es geom\'{e}tricas; por isso Kant n\~{a}o compartilharia o projeto de Galileu de matematiza\c{c}\~{a}o da F\'{\i}sica.\footnote{Caruso \& Moreira, \textit{op. cit.}. Os autores afirmam que o matem\'{a}tico Bernhard Riemann recuperar\'{a} a possibilidade de se pensar o espa\c{c}o como fen\^{o}meno f\'{\i}sico e geom\'{e}trico.}  De qualquer forma, podemos antecipar aqui que essa discuss\~{a}o sobre ser o espa\c{c}o objeto da F\'{\i}sica e/ou da Geometria ser\'{a} superada, ou melhor, dispens\'{a}vel, na Cr\'{\i}tica da Raz\~{a}o Pura, uma vez que nesta obra o espa\c{c}o ser\'{a} pensado de forma original, diferente de sua concep\c{c}\~{a}o no per\'{\i}odo pr\'{e}-cr\'{\i}tico. Esse problema aparece com Newton, e \'{e} algo em rela\c{c}\~{a}o a que Kant ir\'{a} se posicionar na \textit{Cr\'{\i}tica da Raz\~{a}o Pura} -- as ci\^{e}ncias F\'{\i}sica e Matem\'{a}tica, que lidam com a experi\^{e}ncia poss\'{\i}vel, a qual pode ser conhecida, distinguem-se da Metaf\'{\i}sica, que se ocupa com ideias que n\~{a}o podemos perceber, como ``Deus'', ``imortalidade da alma'' e ``liberdade''.

Ao se tomar o espa\c{c}o como absoluto n\~{a}o se est\'{a} na verdade legitimando-o como objeto da F\'{\i}sica; \'{e} preciso reconhecer o quanto soa estranho a proposi\c{c}\~{a}o de tal conceito sem experiment\'{a}-lo e sem registr\'{a}-lo matematicamente, constituindo-se apenas em algo logicamente sugerido, o que em nada garante sua realidade ontol\'{o}gica. E pode ser por ter percebido n\~{a}o haver conex\~{a}o necess\'{a}ria na realidade entre extens\~{a}o -- por intera\c{c}\~{a}o ou movimento de corpos -- e espa\c{c}o que Kant se limita a argumentar acerca da dimensionalidade da extens\~{a}o em seu escrito pr\'{e}-cr\'{\i}tico.    
Quanto ao que o jovem Kant adota do programa newtoniano, vemos que o espa\c{c}o -- ou mais precisamente a extens\~{a}o -- \'{e} pensado a partir das leis de for\c{c}a, vislumbrando-se a extens\~{a}o na disposi\c{c}\~{a}o dos corpos ao se compor uma ordem, a qual \'{e} provocada pela for\c{c}a -- em \'{u}ltima an\'{a}lise \'{e} como se a extens\~{a}o dependesse da for\c{c}a, e esta \'{e} calculada pela dist\^{a}ncia entre os corpos; e uma vez que a medi\c{c}\~{a}o deste fato f\'{\i}sico \'{e} um processo racional, \'{e} a mente humana que passa a ser legitimadora do conhecimento, e n\~{a}o a gra\c{c}a ou o desejo divinos -- ali\'{a}s, \'{e} tal enfoque que consistir\'{a} a revolu\c{c}\~{a}o copernicana \`{a}s avessas como projeto da \textit{Cr\'{\i}tica da Raz\~{a}o Pura}, na medida em que Kant pretende com isso explicitar o nosso modo de conhecer por condi\c{c}\~{o}es mentais a priori, necess\'{a}rias e universais, projeto esse que ele denomina de \textit{Filosofia Transcendental}; neste projeto cr\'{\i}tico, ser\~{a}o descritos as possibilidades e os limites da mente ou raz\~{a}o humana; e conquanto uma tend\^{e}ncia natural da raz\~{a}o, a metaf\'{\i}sica extrapola o campo da experi\^{e}ncia poss\'{\i}vel. Naturalmente, este novo espa\c{c}o a priori n\~{a}o pode ter qualquer de suas propriedades determinadas por uma lei da F\'{\i}sica, produto do intelecto humano, como imaginou o jovem Kant.

\section{A natureza do espa\c{c}o}

N\~{a}o podemos perder de vista que, em 1747, Kant abre m\~{a}o de provar a tridimensionalidade do ``espa\c{c}o'', limitando-se a discutir a da ``extens\~{a}o''; podemos reconhecer que ele n\~{a}o faz uso do conceito de ``espa\c{c}o absoluto'' de Newton; ele parte do conceito de for\c{c}a para fundamentar a ideia de extens\~{a}o; precisamos notar tamb\'{e}m que uma vez poss\'{\i}vel ou pertinente um c\'{a}lculo a partir de uma lei de for\c{c}a, esta no\c{c}\~{a}o fundamental dos eventos da natureza n\~{a}o assume uma conota\c{c}\~{a}o metaf\'{\i}sica, consolidando sua plausibilidade em eventos emp\'{\i}ricos ilustrativos da express\~{a}o da lei gravitacional. Assim, a extens\~{a}o pensada matematicamente em seu sistema de for\c{c}as e disposi\c{c}\~{a}o dos corpos \'{e} a revela\c{c}\~{a}o emp\'{\i}rica que pode ser objeto da F\'{\i}sica. Contudo, o espa\c{c}o absoluto n\~{a}o pode ser apreendido empiricamente, apenas o espa\c{c}o relacional, \textit{i.e.}, aquele entendido como extens\~{a}o nas configura\c{c}\~{o}es da lei de for\c{c}a, como concebe o jovem Kant. Cabe aqui, para marcar diferen\c{c}a com essa perspectiva de Kant, citar como Max Jammer refere-se ao entendimento de Newton sobre os espa\c{c}os relativo e absoluto:\footnote{Max Jammer, \textit{Conceito de Espa\c{c}o: A Hist\'{o}ria das Teorias do Espa\c{c}o na F\'{\i}sica}. Rio de Ja\-nei\-ro: Contraponto, 2010, p.~135.}

\begin{quotation}
\noindent \textit{``Ao acreditar que tempo, espa\c{c}o, lugar e movimento eram conceitos bem conhecidos por todos, Newton, como vemos, n\~{a}o se sentiu convocado a fornecer uma defini\c{c}\~{a}o rigorosa e precisa desses termos. Todavia, como essas no\c{c}\~{o}es s\'{o} surgiam ligadas a objetos sens\'{\i}veis, certos preconceitos aderiam a elas. Para super\'{a}-los, Newton julgou necess\'{a}rio estabelecer distin\c{c}\~{o}es entre o absoluto e o relativo, o verdadeiro e o aparente, o matem\'{a}tico e o comum. Visto que o espa\c{c}o era homog\^{e}neo e indiferenciado, suas partes eram impercept\'{\i}veis e indistingu\'{\i}veis para nossos sentidos, de modo que era preciso substitu\'{\i}-lo por medidas sens\'{\i}veis. Esses sistemas de coordenadas, como hoje os chamamos, s\~{a}o os espa\c{c}os relativos de Newton.''}
\end{quotation}

E continua:

\begin{quotation}    
\noindent \textit{``Na f\'{\i}sica moderna, os sistemas de coordenadas n\~{a}o passam de uma fic\c{c}\~{a}o \'{u}til. Mas n\~{a}o era assim para Newton. Dada a concep\c{c}\~{a}o newtoniana realista dos objetos matem\'{a}ticos, \'{e} f\'{a}cil compreender por que esses espa\c{c}os relativos formavam `medidas sens\'{\i}veis'. N\~{a}o s\'{o} o corpo que servia de refer\^{e}ncia era acess\'{\i}vel aos nossos sentidos, como o `espa\c{c}o relativo' dependia dele. Mas essa acessibilidade \`{a} percep\c{c}\~{a}o sensorial produzia uma no\c{c}\~{a}o que s\'{o} tinha validade provis\'{o}ria e \`{a} qual faltava generalidade. Era bem poss\'{\i}vel que n\~{a}o houvesse nenhum corpo em repouso, ao qual os lugares e movimentos dos outros corpos pudessem ser referidos; em suma, todos aqueles espa\c{c}os relativos talvez fossem sistemas de coordenadas em movimento. Mas movendo-se em qu\^{e}? Para responder a essa pergunta, Newton abandonou o \^{a}mbito da experi\^{e}ncia, ao menos provisoriamente. Com palavras que se tornaram famosas -- `Nas investiga\c{c}\~{o}es filos\'{o}ficas, devemos nos abstrair de nossos sentidos' -- introduziu o espa\c{c}o absoluto e imut\'{a}vel, do qual o espa\c{c}o relativo era apenas uma medida. O grau \'{u}ltimo de exatid\~{a}o, a verdade suprema, s\'{o} poderia ser alcan\c{c}ado em refer\^{e}ncia a esse espa\c{c}o absoluto.'\footnote{Max Jammer, \textit{op. cit.}, p.~136.}}
\end{quotation}

E Jammer ainda chama aten\c{c}\~{a}o para algumas quest\~{o}es pertinentes: 

\begin{quotation}
\noindent \textit{``Nesse ponto podemos indagar: o que garantiria a verdade \'{u}ltima do espa\c{c}o absoluto (...). Na \'{e}poca de Newton, essa quest\~{a}o tornou-se sumamente controvertida e assim permaneceu at\'{e} o come\c{c}o do s\'{e}culo XX. Seria o conceito de espa\c{c}o absoluto uma necessidade da F\'{\i}sica? Ou seria poss\'{\i}vel construir um sistema conceitual coerente que explicasse todos os fen\^{o}menos f\'{\i}sicos sem recorrer a esse conceito?''\footnote{\textit{Ibid},  p.~136-137.}}
\end{quotation}

Pelo que vimos sobre a argumenta\c{c}\~{a}o do jovem Kant a respeito do problema da dimensionalidade do espa\c{c}o, ele parece estar ciente dessa controv\'{e}rsia e toma uma posi\c{c}\~{a}o que n\~{a}o contempla a ideia de espa\c{c}o absoluto naquela sua argumenta\c{c}\~{a}o, fazendo predominar uma explica\c{c}\~{a}o dos fen\^{o}menos f\'{\i}sicos o mais emp\'{\i}rica poss\'{\i}vel, por assim dizer, ``pr\'{o}xima'' de nossa percep\c{c}\~{a}o ou intui\c{c}\~{a}o sens\'{\i}vel.

Na CRP, contudo, o espa\c{c}o assumir\'{a} uma conota\c{c}\~{a}o transcendental, como condi\c{c}\~{a}o de possibilidade a priori da intui\c{c}\~{a}o, apreens\~{a}o sens\'{\i}vel dos dados f\'{\i}sicos; espa\c{c}o assim consiste em um modo como podemos conhecer, sendo uma forma pura para percebermos os fen\^{o}menos, os dados sens\'{\i}veis.\footnote{No seu texto pr\'{e}-cr\'{\i}tico, Kant vislumbrou a possibilidade de o espa\c{c}o constituir-se de diferentes n\'{u}meros de dimens\~{o}es. Tal perspectiva ser\'{a} pertinente a partir da CRP: se espa\c{c}o \'{e} condi\c{c}\~{a}o pura da experi\^{e}ncia, e se esta pode ser pensada por leis f\'{\i}sicas e medi\c{c}\~{o}es matem\'{a}ticas executadas sobre os fen\^{o}menos percebidos, ent\~{a}o a matem\'{a}tica como ci\^{e}ncia poss\'{\i}vel, assim como a F\'{\i}sica, pode construir conceitos diversos, como, por exemplo, o de ``dimens\~{a}o''. Sendo assim, espa\c{c}o n\~{a}o \'{e} fundamentalmente tridimensional -- ele o \'{e} na Geometria Euclidiana, e conforme nossa percep\c{c}\~{a}o cotidiana --, mas antes \'{e} forma a priori da sensibilidade, condi\c{c}\~{a}o para se perceber os dados emp\'{\i}ricos, n\~{a}o \'{e} um dado natural, nem absoluto ou particular, \'{e} forma pura, \'{e} condi\c{c}\~{a}o mental para percep\c{c}\~{a}o da natureza.}

Caruso e Moreira\footnote{\textit{Op. cit.}}  entendem que uma das contribui\c{c}\~{o}es do jovem Kant \'{e} a de romper com a vis\~{a}o aristot\'{e}lica sobre o espa\c{c}o, pensando for\c{c}a como causa eficiente deste por meio do conceito de ordem. Aqui vale uma digress\~{a}o. A partir disso pode-se reconhecer a pertin\^{e}ncia da ideia de espa\c{c}o curvo pensada por Einstein, se se admite que a possibilidade dessa curvatura deve-se a alguma for\c{c}a exercida sobre o espa\c{c}o.

O espa\c{c}o absoluto \'{e} defendido por Newton, um espa\c{c}o metaf\'{\i}sico, com pertin\^{e}ncia l\'{o}gica at\'{e} -- ou mesmo necessidade l\'{o}gica, uma vez que em tese \'{e} de tal modo oposto \`{a} compreens\~{a}o de espa\c{c}o relacional de Leibniz --, mas n\~{a}o propriamente f\'{\i}sico, ainda que Newton o considere uma verdade ontol\'{o}gica.  Na perspectiva de Newton, o espa\c{c}o absoluto permite, por sua necessidade l\'{o}gica e ontol\'{o}gica, apreens\~{a}o, descri\c{c}\~{a}o, medi\c{c}\~{a}o dos fen\^{o}menos que ocorrem em sua abrang\^{e}ncia. \'{E} o que podemos ver brevemente em palavras de Max Jammer:\footnote{Max Jammer, \textit{op. cit.}, 
p.~137.}

\begin{quotation}
\noindent \textit{``Para Newton, o espa\c{c}o absoluto era uma necessidade l\'{o}gica e ontol\'{o}gica. Era um pr\'{e}-requisito necess\'{a}rio para a validade da primeira lei do movimento: `Todo corpo preserva o estado de repouso ou de movimento uniforme em linha reta, a menos que seja compelido a modificar esse estado por for\c{c}as imprimidas sobre ele.' O movimento retil\'{\i}neo uniforme exigia um sistema de refer\^{e}ncia diferente do de qualquer espa\c{c}o relativo arbitr\'{a}rio. Al\'{e}m disso, o estado de repouso pressupunha tal espa\c{c}o absoluto.''}
\end{quotation}

E para o Kant no per\'{\i}odo cr\'{\i}tico, o espa\c{c}o n\~{a}o \'{e} f\'{\i}sico, nem \'{e} uma verdade ontol\'{o}gica exterior \`{a} mente humana; o que o esp\'{\i}rito ou a mente apreende s\~{a}o objetos extensos ``espacialmente'', ou melhor, em uma ordena\c{c}\~{a}o, em rela\c{c}\~{o}es; o espa\c{c}o \'{e} uma forma ou condi\c{c}\~{a}o mental para a percep\c{c}\~{a}o das coisas nas rela\c{c}\~{o}es entre si, em uma extens\~{a}o. Ser\'{a} que resqu\'{\i}cios dessa compreens\~{a}o que alcan\c{c}ar\'{a} em sua maturidade podem estar contidos no discreto abandono da tese de que seria poss\'{\i}vel provar a tridimensionalidade do espa\c{c}o e n\~{a}o apenas a da extens\~{a}o, j\'{a} no texto de 1747?
Podemos fazer outra e r\'{a}pida digress\~{a}o ao considerarmos as palavras de Kant, no seu texto de juventude, sobre ser a mente ou ``alma'' humana uma esp\'{e}cie de estado que se modifica com as impress\~{o}es causadas pelos corpos quando percebidos pela mente, \textit{i.e.}, esta \'{e} afetada como se fosse tamb\'{e}m uma ``subst\^{a}ncia'' que sofre os efeitos da mat\'{e}ria do mundo, por assim dizer: ``(...) \textit{a mat\'{e}ria, por interm\'{e}dio da for\c{c}a, que tem no movimento, altera o estado da alma por quanto a alma representa o pr\'{o}prio mundo}.''\footnote{Kant, \textit{op. cit.}, p.~7-8.}  E \'{e} curioso que, na CRP, Kant pretende elucidar como os dados naturais nos aparecem, como o mundo \'{e} representado pela mente, descrevendo o ``aparelho de medida mental'' na sua capacidade de ser afetado pelo mundo e enquanto afeta este apreendendo-o como fen\^{o}meno, \textit{i.e.}, como aparece a ele segundo as condi\c{c}\~{o}es de possibilidade a priori do conhecimento da experi\^{e}ncia, condi\c{c}\~{o}es puras que s\~{a}o mentais.

\section{Coment\'{a}rios finais}
Retomemos a exposi\c{c}\~{a}o sobre a argumenta\c{c}\~{a}o de Kant acerca da natureza do espa\c{c}o. O jovem Kant entende o espa\c{c}o como objeto de estudo da Geometria e n\~{a}o da F\'{\i}sica -- uma vez que n\~{a}o intu\'{\i}do no dom\'{\i}nio da experi\^{e}ncia, e n\~{a}o sendo tamb\'{e}m um pressuposto metaf\'{\i}sico, como concebeu Newton com a ideia de ``espa\c{c}o absoluto''. Portanto, o espa\c{c}o passaria a ser abordado com legitimidade unicamente pela Matem\'{a}tica. Este tipo de quest\~{a}o vai voltar a ser debatido, tanto cient\'{\i}fica como filosoficamente; por exemplo, quando Bernhard Riemann defende que ``\textit{a matem\'{a}tica poderia determinar a estrutura m\'{e}trica do espa\c{c}o.}''. Tal afirmativa antecipa as correla\c{c}\~{o}es entre subst\^{a}ncia, espa\c{c}o f\'{\i}sico e espa\c{c}o geom\'{e}trico que est\~{a}o por tr\'{a}s dos Fundamentos da Teoria da Relatividade Geral de Einstein e, em \'{u}ltima an\'{a}lise, fazem Einstein aceitar a exist\^{e}ncia de um \'{e}ter grav\'{\i}fico\footnote{Albert Einstein, O \'{e}ter e a teoria da relatividade. Confer\^{e}ncia feita na Universidade de Leyden, em maio de 1920, e publicada em franc\^{e}s em \textit{R\'{e}flexions sur l'Electrodynamique, l'\'{E}hter, la G\'{e}ometrie e la Relativit\'{e}}, Paris: Gauthier-Villars, 1972.} embora descartasse o \'{e}ter lumin\'{\i}fero.

Concluindo, gostar\'{\i}amos de lembrar que n\~{a}o h\'{a} outro lugar na obra de Kant no qual ele retorne a tentar estabelecer uma base f\'{\i}sica para a dimensionalidade do espa\c{c}o. Apenas nos manuscritos coligidos nos \textit{Opus Postumum}\footnote{E. Kant, \textit{Opus Postumum -- passage des principes m\'{e}taphysiques de la science de la nature \`{a} la physique}, translation, presentation and notes by F. Marty, Paris, Press Univ. de France, 1986, p.~131.}  h\'{a} uma refer\^{e}ncia a esta quest\~{a}o, mas que, quis o destino, em nada colabora para esclarecer como o Kant maduro refletiu sobre seus escritos da juventude. \'{E} com esse fragmento, e chamando aten\c{c}\~{a}o para o ponto no qual ele se interrompe, que gostar\'{\i}amos de concluir o artigo:
\begin{quotation}
\noindent	\textit{``A qualidade do espa\c{c}o e do tempo, por exemplo, que o primeiro tem tr\^{e}s dimens\~{o}es e que o segundo, somente uma, que a revolu\c{c}\~{a}o \'{e} regida pelo quadrado das dist\^{a}ncias s\~{a}o princ\'{\i}pios que ...''}.
\end{quotation}

\end{document}